\documentclass[10pt,conference]{IEEEtran}
\IEEEoverridecommandlockouts

\usepackage{cite}
\usepackage{amsmath,amssymb,amsfonts}
\usepackage{graphicx}
\usepackage{bm}
\usepackage{booktabs} 
\usepackage{diagbox} 
\usepackage{enumitem}
\usepackage{multicol}
\usepackage{multirow}
\usepackage{makecell}
\usepackage{pifont}
\usepackage{subfigure}
\usepackage{xcolor}
\usepackage[linesnumbered,lined,ruled,commentsnumbered]{algorithm2e}
\usepackage{url,hyperref,cleveref}
\usepackage{makecell}

\def\BibTeX{{\rm B\kern-.05em{\sc i\kern-.025em b}\kern-.08em
    T\kern-.1667em\lower.7ex\hbox{E}\kern-.125emX}}

\begin{document}
\title{Low-Resource Domain Adaptation for Speech LLMs via Text-Only Fine-Tuning}



\author{\IEEEauthorblockN{Yangui Fang$^{1,4*}$\thanks{* Yangui Fang and Jing Peng contributed equally to this work.}, Jing Peng$^{2,3*}$, Xu Li$^4$ , Yu Xi$^{2,3}$, Chengwei Zhang$^1$$^{\dagger}$,Guohui Zhong$^1$, Kai Yu$^{2,3}$ \thanks {$^{\dagger}$ Corresponding Author.}}
\IEEEauthorblockA{\textit{$^1$Huazhong University of Science and Technology, School of Electronic Information and Communications}}
\IEEEauthorblockA{\textit{$^2$MoE Key Lab of Artificial Intelligence, AI Institute, X-LANCE Lab, Shanghai Jiao Tong University, Shanghai, China}}
\IEEEauthorblockA{\textit{$^3$Jiangsu Key Lab of Language Computing, Suzhou, China} \textit{{$^4$AISpeech Co., Ltd., Suzhou, China}}}
\IEEEauthorblockA{
 \{fangyg,zhangcw,zhonggh\}@hust.edu.cn~~~danieljingpeng@gmail.com~~~\{yuxi.cs,kai.yu\}@sjtu.edu.cn~~~ xu.li@aispeech.com~~~ 
}
}

\maketitle


\begin{abstract}

Recent advances in automatic speech recognition (ASR) have combined speech encoders with large language models (LLMs) through projection, forming Speech LLMs with strong performance. However, adapting them to new domains remains challenging, especially in low-resource settings where paired speech-text data is scarce. We propose a text-only fine-tuning strategy for Speech LLMs using unpaired target-domain text without requiring additional audio. To preserve speech-text alignment, we introduce a real-time evaluation mechanism during fine-tuning. This enables effective domain adaptation while maintaining source-domain performance. Experiments on LibriSpeech, SlideSpeech, and Medical datasets show that our method achieves competitive recognition performance, with minimal degradation compared to full audio-text fine-tuning. It also improves generalization to new domains without catastrophic forgetting, highlighting the potential of text-only fine-tuning for low-resource domain adaptation of ASR.

\end{abstract}

\section{Introduction}
Automatic speech recognition (ASR) is a core technology in intelligent speech systems. It has evolved from traditional ASR models \cite{levinson1983introduction,lee1989speaker,9536732} to end-to-end (E2E) models \cite{abdel2014convolutional,gulati2020conformer,gao2022paraformer,graves2005framewise,ren2022improving}
and is now undergoing a third paradigm shift toward Speech LLMs \cite{peng2025surveyspeechlargelanguage, arora2025landscapespokenlanguagemodels}. Speech LLMs extend general-purpose LLMs by incorporating audio encoders that extract acoustic features, which are then projected into the LLM input space and combined with text token embeddings~\cite{ma2024embarrassingly,tang2023salmonn}. Empowered by the strong contextual reasoning capabilities of LLMs, Speech LLMs have demonstrated remarkable performance in the field of ASR. Recent models such as SeedASR \cite{bai2024seed} and FireRedASR \cite{xu2025fireredasr} have achieved state-of-the-art results on benchmark datasets including LibriSpeech \cite{panayotovLibrispeechASRCorpus2015} and AISHELL \cite{buAISHELL1OpensourceMandarin2017} , significantly surpassing the performance of previous approaches.

Leveraging the strong learning capacity of Speech LLMs, these models have achieved outstanding recognition performance on widely-used benchmark datasets. However, this success also introduces a significant risk: the training and performance improvements of Speech LLMs heavily depend on the availability of large-scale data. In scenarios where training data are sparse, the model is prone to overfitting to specific domains\cite{kumar2024performance}, resulting in degraded generalization to unseen or broader domains. And a more important question is, in low-resource speech domains, the scarcity of paired audio-text data\cite{singhal2023domain,ashok2024domain} poses a major bottleneck, making it difficult to perform domain-specific fine-tuning and leading to poor recognition accuracy in those domains.

One approach leverages the relative abundance of textual data in low-resource domains. Given sufficient in-domain text, an LLM can generate domain-specific content, and a text-to-speech (TTS) system can synthesize corresponding audio to create pseudo audio-text pairs. However, synthetic speech often lacks naturalness and variability~\cite{dunbar2019zero,10389722}, and the overall pipeline incurs significant computational and engineering costs. Another direction reduces the number of trainable parameters in Speech LLMs\cite{li2023prompting,liao2023zero}, such as prompt tuning or prefix tuning. While these methods lessen retraining overhead, they require domain-specific prompt design, limiting generalizability and introducing additional manual effort.

To address these challenges, we propose a novel training strategy that fine-tunes the Speech LLM using unpaired text data without requiring additional audio, while maintaining cross-modal alignment through real-time alignment evaluation during fine-tuning. Our main contributions are summarized as follows:
\begin{itemize}
    \item To the best of our knowledge, this is the first work in the field of Speech LLMs that explores domain adaptation by fine-tuning the LLM using only text data from the target domain.
    
    \item To mitigate the risk of degradation in speech recognition performance caused by excessive text fine-tuning, we propose a real-time evaluation strategy that employs text data for training while using paired speech-text data for evaluation.
    
    \item Our experiments demonstrate that text-based fine-tuning not only facilitates domain adaptation in low-resource scenarios but also preserves superior recognition performance in the original domain compared to conventional speech-based fine-tuning methods.
\end{itemize}

\section{Related Works}

This section first reviews traditional domain adaptation methodologies and end-to-end speech recognition adaptation strategies, which serve as foundational inspirations for our approach. It then examines solutions based on TTS for addressing domain adaptation in low-resource scenarios, highlighting their limitations. Finally, recent advances in Speech LLMs are discussed as a promising yet still challenging direction, forming the primary focus of this work.

\subsection{Domain Adaptation for Traditional Models}  
Traditional ASR systems typically consist of three independent components: an acoustic model (AM), a pronunciation lexicon (PM), and a language model (LM). The recognition process can be formulated as
\begin{equation}
P_{\mathrm{LM}}(T) \cdot P_{\mathrm{AM}}(T \mid A),
\end{equation}
where \(A\) denotes the input audio signal and \(T\) denotes the corresponding output text sequence.

In domain adaptation, the objective is to leverage a model trained on a source domain to recognize speech from a different target domain. The recognition process under domain adaptation can be formulated as
\begin{equation}
P_{\mathrm{LM}_\phi}(T_\tau) \cdot P_{\mathrm{AM}_\phi}(T_\tau \mid A_\tau),
\end{equation}
where \(\phi\) denotes the model parameters trained on the source domain, \(\tau\) denotes the target domain, \(A_\tau\) denotes the input audio from the target domain, and \(T_\tau\) denotes the corresponding output text sequence.

Domain mismatches between training and target data may cause significant performance degradation. To mitigate this issue, it is often necessary to adapt the acoustic model or the language model trained on the source domain. Given the lower retraining cost, language models are typically adapted more frequently. Specifically, integrating domain-specific language models facilitates effective adaptation of traditional systems. And the training process can be formulated as
\begin{equation}
\label{eq:domain_adaptation}
\resizebox{0.91\hsize}{!}{
  $P_{\mathrm{LM}_\phi}(T_\tau) \cdot P_{\mathrm{AM}_\phi}(T_\tau \mid A_\tau) \longrightarrow P_{\mathrm{LM}_{\phi+\tau}}(T_\tau) \cdot P_{\mathrm{AM}_\phi}(T_\tau \mid A_\tau)$,
}
\end{equation}
where \(P_{\mathrm{LM}_{\phi+\tau}}\) denotes the language model originally trained on the source domain and adapted to the target domain \(\tau\), while \(P_{\mathrm{AM}_\phi}\) remains fixed with the original parameters \(\phi\). The proposed method draws upon the conventional strategy.

\subsection{Domain Adaptation For End to End Models}
In end-to-end (E2E) models, domain adaptation can also be performed by replacing the LM; however, the LM is no longer a necessary component of E2E architectures. In such systems, the LM is typically incorporated via shallow fusion~\cite{zhao2019shallow}, deep fusion~\cite{thomas2022integrating}, or cold fusion~\cite{sriram2017cold}. For deep fusion and cold fusion approaches, domain adaptation through LM replacement is challenging because the LM is implicitly integrated into the model. The recognition probability in this case can be formulated as
\begin{equation}
\label{eq:e2e_implicit_lm}
P_{\mathrm{E2E}_\phi}\left(T_\tau \mid A_\tau, \mathrm{LM}_\phi^{\text{(implicit)}}\right),
\end{equation}
where \(\mathrm{LM}_\phi^{\text{(implicit)}}\) indicates that the LM is implicitly integrated into the model with parameters \(\phi\).

Another approach is to augment the model with relevant information from the target domain~\cite{aleksic2015bringing}. For example, Pundak et al.~\cite{pundak2018deep} incorporate contextual information through a context encoder and an attention mechanism. Huang et al.~\cite{huang2023contextualized} propose combining context embeddings with acoustic representations extracted from the acoustic encoder and text representations obtained using a prediction network based on previous sub-word units. Yang et al.~\cite{yang2023two} further propose an effective method to construct high-quality context lists for unified streaming and non-streaming E2E models. However, these methods require additional resources or supervision.

\subsection{Domain Adaptation For Synthetic Speech}


The development of synthetic speech technology enables the use of speech synthesized from text in data-scarce domains to promote domain adaptation. Su et al.~\cite{su2024corpus} studied an E2E pipeline where LLMs are used to generate text, followed by speech synthesis and domain adaptation, potentially eliminating the reliance on textual data. D et al.~\cite{d2023can} adopt the method proposed by Thomas et al.~\cite{thomas2022integrating} to investigate whether synthetic speech can be substituted by injected contextual information. 

However, a major limitation of the synthetic speech approach is that it requires substantial computational resources for speech synthesis, and its effectiveness heavily depends on the quality of the generated speech. The objective is to ensure that the model performance on synthetic speech approximates that on real speech, which can be formalized as
\begin{equation}
\label{eq:synthetic_speech_adaptation}
P_{\mathrm{E2E}_\phi}\left(T_\tau \mid A_{\tau}^{\text{(synth)}}\right) \approx P_{\mathrm{E2E}_\phi}\left(T_\tau \mid A_\tau\right),
\end{equation}
where \(T_\tau\) denotes the output text sequence for the target domain, \(A_\tau\) denotes the input audio from the target domain, and \(A_{\tau}^{\text{(synth)}}\) represents the synthetic audio generated from text for the target domain.

\subsection{Domain Adaptation For Speech LLMs}
Speech large language models (Speech LLMs) leverage extensive general-domain textual information acquired during pre-training and can further inject domain-specific knowledge by exploiting their contextual capabilities. Yang et al.~\cite{yang2024mala} improve recognition performance by prompting the LLM with hot words extracted from slide text. Similarly, Lakomkin et al.~\cite{lakomkin2024end} incorporate metadata such as titles to enhance recognition accuracy. Nevertheless, obtaining specific contextual information for each speech utterance remains challenging. 

Li et al.~\cite{li2023prompting} propose using domain-specific text prompts to improve recognition performance in a particular domain; however, the prompts must be carefully designed, and devising a stable and broadly applicable prompting strategy remains an open problem. Liao et al.~\cite{liao2023zero} investigate using LLMs to generate descriptions for corresponding speech segments, allowing models like Whisper to leverage these descriptions for recognition enhancement. Despite its potential, generating a text description for each utterance is often impractical.
\begin{figure*}[!ht]
\centering
\includegraphics[width=0.99\textwidth]{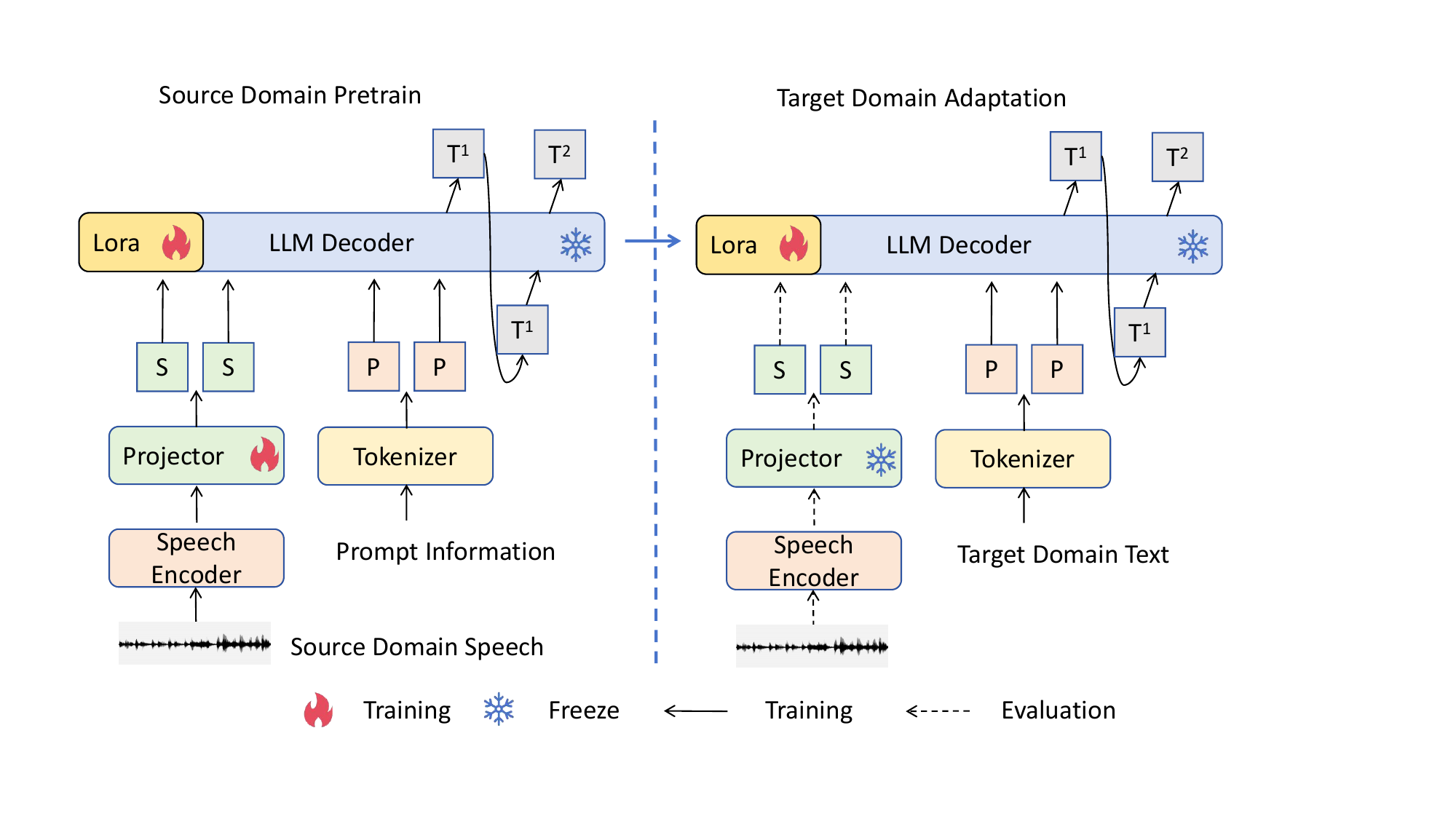}
\caption{An overview of our two-stage training framework. Left: Source domain pretraining aims to achieve cross-modal alignment between speech and text, following mainstream training strategies. Right: Target domain adaptation seeks to maintain alignment while improving performance on the target domain. During training, the LLM is fine-tuned with LoRA using text-only data, while real-time evaluation of alignment is conducted using text-audio paired data.}
\label{Fig:Target Domain Text Fine-tune LLM Adaptation}
\end{figure*}
\section{Domain Adaptation for Speech LLMs via Text-Only Fine-Tuning}

\subsection{Fine-Tuning LLM on Target Domain Text for Adaptation}

To construct a Speech LLM, a general-purpose LLM is connected to a pretrained speech encoder via an adapter module. The resulting architecture is trained on paired audio-text data from a source domain to achieve cross-modal alignment between speech and text. The recognition process can be formalized as
\begin{equation}
\label{eq:source_pretrain}
\resizebox{0.91\hsize}{!}{
$P_{\mathrm{SpeechLLM}}\left(
T_{\tau} \,\middle|\,
\underbrace{\mathrm{Projector}_{\phi}\left(\mathrm{Encoder}_{\phi}(A_{\tau})\right)}_{\text{Acoustic feature mapping}},\ 
\mathrm{LLM}_{\theta_s + \phi}
\right)$,
}
\end{equation}
where \(\mathrm{LLM}_{\theta_s + \phi}\) denotes a large language model initially pre-trained on general-domain text \(\theta_s\) and subsequently adapted to the source domain via cross-modal training with paired speech-text data. The model parameters, including the pretrained encoder \(\mathrm{Encoder}_{\phi}\), the projection module \(\mathrm{Projector}_{\phi}\), and the source-domain adapted LLM \(\mathrm{LLM}_{\theta_s + \phi}\) , are jointly optimized by minimizing the cross-entropy loss between the predicted transcription and the ground-truth text. We refer to this stage as source domain pretraining, as illustrated on the left side of Fig.~\ref{Fig:Target Domain Text Fine-tune LLM Adaptation}. Both the projector and the LLM are included in the training process to achieve better generalization ability~\cite{kumar2024performance}.

While traditional domain adaptation methods typically perform end-to-end fine-tuning on speech-text paired data, if this approach is directly applied to Speech LLMs, the speech recognition process will be formulated as
\begin{equation}
\label{eq:e2e_adapt}
\resizebox{0.91\hsize}{!}{
$P_{\mathrm{SpeechLLM}}\left(
T_{\tau} \,\middle|\,
\mathrm{Projector}_{\phi+\tau}\left(\mathrm{Encoder}_{\phi+\tau}(A_{\tau})\right),\ 
\mathrm{LLM}_{\theta_s + \phi + \tau}
\right)$,
}
\end{equation}
where \(\mathrm{LLM}_{\theta_s + \phi + \tau}\) denotes the large language model initially pre-trained on general-domain text, adapted to the source domain via cross-modal training, and further fine-tuned on the target domain speech-text data for domain-specific adaptation. However, this approach faces two significant challenges: (1) the Speech LLMs are prone to overfitting when limited domain data is available, (2) the training process acquires sufficient speech-text paired corpora in low-resource domains, such as those with specialized terminology, is difficult.

To address these challenges, we propose a hierarchical adaptation strategy, as shown on the right side of Fig.~\ref{Fig:Target Domain Text Fine-tune LLM Adaptation}. In the target domain adaptation phase, only the LoRA layers within the LLM decoder are fine-tuned by leveraging only text data. The training objective is the standard autoregressive language modeling loss, which minimizes the negative log-likelihood of the target domain text sequences.

After adaptation, the speech recognition process can be formulated as
\begin{equation}
\label{eq:text_only_adapt}
\resizebox{0.91\hsize}{!}{
$P_{\mathrm{SpeechLLM}}\left(
T_{\tau} \,\middle|\,
\mathrm{Projector}_{\phi}\left(\mathrm{Encoder}_{\phi}(A_{\tau})\right),\ 
\mathrm{LLM}_{\theta_s + \phi + \tau_{\text{text}}}
\right)$,
}
\end{equation}
where \(\mathrm{LLM}_{\theta_s + \phi + \tau_{\text{text}}}\) denotes the large language model initially pre-trained on general-domain text, adapted to the source domain via cross-modal training, and further fine-tuned on target domain text data using LoRA-based lightweight adaptation.

This adaptation strategy offers three advantages: (a) it preserves the original-domain generalization capability of the acoustic feature extractor, including the projector and encoder; (b) it injects domain-specific knowledge by leveraging the relatively abundant textual data; and (c) it mitigates the risk of overfitting arising from joint optimization of acoustic and textual modalities.

\subsection{Real Time Evaluation}

The LoRA adapters in LLMs exhibit dual functionalities across two stages in our method: \textit{cross-modal alignment optimization} (enhancing speech-to-text representation mapping during the initial source domain pretraining stage) and \textit{language modeling enhancement} (learning domain-specific linguistic patterns through target domain text adaptation in the second stage). 

However, while conventional text-only fine-tuning approaches typically focus exclusively on improving language modeling capability, monitored through the language modeling loss \(\mathcal{L}_{\text{LM}}\), they often neglect the preservation of speech-text alignment established during the initial cross-modal pretraining. This omission can lead to degradation in alignment performance, as the model overfits to the text modality and loses its ability to effectively integrate acoustic representations.

To address this limitation, we propose a dynamic evaluation strategy that ensures the cross-modal alignment is maintained throughout the text adaptation process:
\begin{itemize}
    \item \textbf{Training:} Update the LoRA parameters of the LLM by minimizing the language modeling loss \(\mathcal{L}_{\text{LM}}\).
    \item \textbf{Evaluation:} Freeze the encoder, projector, and base LLM parameters, and integrate the updated LoRA parameters to evaluate the speech recognition loss on paired speech-text data.
\end{itemize}

By continuously monitoring the speech-text alignment of the text-adapted model during fine-tuning, we ensure that text-only adaptation enhances domain-specific language modeling while preserving robust cross-modal representation capabilities.





\begin{table*}[]
    \centering
    \caption{Details of the selected datasets and characteristics of each dataset}
    \label{tab:dataset}
    \renewcommand{\arraystretch}{1.3} 
    \newcolumntype{S}{>{\small}c} 
    \begin{tabular}{c|S|S|S|S|S|S}
      \toprule 
      \textbf{Dataset} & 
      \textbf{Train} & 
      \textbf{Dev} & 
      \textbf{Test} & 
      \textbf{Hours} & 
      \textbf{Source} & 
      \textbf{Features} \\
      \midrule 
      Librispeech & 181,240 & 5,567 & 5,558 & 1,000 & Audiobook & Source domain dataset \\
      Medical & 381 & 385 & 5,895 & 8 & Medical & Small-scale medical domain dataset \\
      Slidespeech & 481,930 & 1,801 & 3,188 & 473 & Online Meeting & Medium-scale conference domain dataset \\
      Gigaspeech & 8,282,988 & 5,715 & 19,930 & 10,000+ & Internet & Large-scale general-purpose multi-domain dataset \\
      \bottomrule 
    \end{tabular}
\end{table*}



\section{Experimental Setup}

\subsection{Dataset}
\label{sub:Dataset}



Table~\ref{tab:dataset} summarizes the datasets used. LibriSpeech~\cite{panayotovLibrispeechASRCorpus2015} (1,000 hours) serves as the source domain, comprising audiobook recordings with test-clean and test-other subsets. For cross-domain evaluation, we adopt SlideSpeech~\cite{wangSlideSpeechLargeScale2024} (470 hours) from the online meeting domain and Medical~\cite{medical} (8 hours) from the healthcare domain, simulating diverse low-resource scenarios across academic and specialized medical fields. Additionally, GigaSpeech~\cite{chenGigaSpeechEvolvingMultidomain2021} (10,000+ hours) provides large-scale evaluation for generalization assessment. These datasets collectively support rigorous cross-domain adaptation experiments.

\begin{table*}[htbp]
\caption{Detailed analysis of substitution, deletion, and insertion errors under different fine-tuning strategies, along with evaluation results on source, target and other domains. The speech LLM is trained on Librispeech.}
\label{tab:anaysys}
\centering
\renewcommand{\arraystretch}{1.4} 

\begin{tabular}{cc|c|c|ccc|cc}
\toprule
  
\multicolumn{2}{c|}{\textbf{Fine-Tune Strategy}} & 
\textbf{Source Domain} & 
\textbf{Target Domain} & 
\multicolumn{3}{c|}{\textbf{Error Types}} & 
\multicolumn{2}{c}{\textbf{Other Domain}} \\
\cmidrule(lr){1-2} \cmidrule(lr){3-3} \cmidrule(lr){4-4} \cmidrule(lr){5-7} \cmidrule(lr){8-9}
 \textbf{Text} & \textbf{Speech} & \textbf{LibriSpeech} & \textbf{SlideSpeech} & \textbf{Sub} & \textbf{Del} & \textbf{Ins} & \textbf{Medical} & \textbf{GigaSpeech} \\
\midrule
 -- & -- &  \textbf{3.73 / 6.52} & 27.59 & 7.63 & 5.38 & 14.58 & 13.03 & 39.47 \\
  SlideSpeech & -- & 4.36 / 6.55 & 16.22 & 6.01 & 8.35 & \textbf{1.66} & \textbf{13.08} & \textbf{17.09} \\ 
  -- & SlideSpeech & 9.71 / 12.47 & 12.99 & 5.43 & 3.13 & 4.42 & 20.69 & 18.13 \\
 SlideSpeech & SlideSpeech & 9.48 / 12.16 & \textbf{12.92} & \textbf{5.39} & \textbf{3.07} & 4.46 & 20.67 & 17.92 \\ 
\midrule
\end{tabular}
\end{table*}

\subsection{Model Architecture}
\label{sub:BaselineModel}
As illustrated in Figure~\ref{Fig:Target Domain Text Fine-tune LLM Adaptation}, our Speech LLM consists of a speech encoder, a projector, and an LLM decoder. We adopt Whisper-large-v3~\cite{radfordRobustSpeechRecognition2023} as the encoder, with 635M parameters and 128-dimensional MFCC features, padding all inputs to 30 seconds. The encoder has a downsampling factor of 2, yielding a fixed output of 1500 frames with 1280 dimensions.

The projector comprises two linear layers with a ReLU activation, totaling 20M parameters, and applies a downsampling factor of 5 by folding every 5 tokens into 1. Remaining tokens are discarded if fewer than five.

For the decoder, we use Qwen2.5-7B-Instruct~\cite{qwen2.5}, a 7B-parameter LLM. We fine-tune it using LoRA, applying modifications to the query, key, value, and output projection matrices in each decoder layer, while freezing the rest. LoRA is configured with a rank of 64, dropout of 5\%, and $\text{LoRA}_\alpha$ of 16, introducing 161M trainable parameters.



\subsection{Training Strategy}
\label{sub:finetune-strategy}
Training uses the Adam optimizer with $\beta_1 = 0.9$, $\beta_2 = 0.98$, weight decay of $10^{-5}$, and a warmup schedule with 1,000 steps to a base learning rate of $10^{-4}$. For text-only fine-tuning, the learning rate is reduced to $5 \times 10^{-6}$ with a 100-step warmup. 

In the source domain pretraining stage, the training proceeds in two phases: first optimizing the projector for 3--5 epochs until convergence, followed by joint optimization of the projector and LoRA parameters for 1--2 epochs, following the main training strategy such as SLAMOON~\cite{geng2025osum} and OSUM~\cite{tang2023salmonn}.

For target domain adaptation, we fine-tune the Speech LLM using three strategies to enable a systematic comparison of their effectiveness.

\begin{enumerate}
    \item Text fine-tuning: We use the text of the training data to fine-tune the LLM in Speech LLM through pre-training. The trainable parameters at this time are the pre-trained LoRA.
    \item Speech fine-tuning: The entire Speech LLM is fine-tuned using speech-text pairs, which is the same as the way to train speech LLM. The trainable parameters at this time are Linear+LoRA.
    \item Text-then-Speech fine-tuning: Take the LoRA parameters of the LLM after text fine-tuning to continue fine-tuning  of Speech LLM using the speech-text pair.
\end{enumerate}

\begin{table}[htbp]
\renewcommand{\arraystretch}{1.3} 
\caption{Cross-domain speech recognition text fine-tuning results. The model is \textbf{Speech LLM (8B)} with \textbf{181M} trainable parameters, trained on \textbf{LibriSpeech}. Fine-tuning datasets include Medical (less than 1h), SlideSpeech (473h), and Gigaspeech (10k+h). WER(\%) is reported for both source and target domains.}
\centering
\setlength{\tabcolsep}{3pt} 
\begin{tabular}{c|cc|ccc}
\toprule
\textbf{\makecell[c]{Text\\Fine-tune Data}} & 
\multicolumn{2}{c|}{\textbf{\makecell{Source Domain\\ (WER\%)}}} & 
\multicolumn{3}{c}{\textbf{\makecell{Target Domain\\(WER\%)}}} \\
\cmidrule(lr){2-3} \cmidrule(l){4-6}
& \textbf{Clean} & \textbf{Other} & 
\textbf{Medical} & \textbf{SlideSpeech} & \textbf{Gigaspeech} \\
\midrule
- & \textbf{3.73} & \textbf{6.52} & 13.03 & 27.59 & 39.47 \\
Medical  & 4.05 & 5.97 & 12.48 & 16.53 & 18.04 \\
SlideSpeech & 4.36 & 6.55 & 13.08 & \textbf{16.22} & \textbf{17.09} \\
Gigaspeech & 4.90 & 6.55 & \textbf{12.26} & 17.09 & 17.79 \\
\bottomrule
\end{tabular}
\label{tab:effective}
\end{table}

\section{Evaluation and Results Analysis}
\subsection{Generalization Evaluation of Speech LLM After Source Domain Pretraining}
\label{sub:Generalization}
As shown in the first row of Table~\ref{tab:effective}, the base model is trained on LibriSpeech as the source domain. Consistent with findings in \cite{kumar2024performance}, the Speech LLM achieves relatively good performance on LibriSpeech due to its large parameter size and superior learning capacity. (Note that with more training epochs, the performance could be further improved on source dataset; however, we limit the epochs for efficient experimentation and relatively good generalization ability.) However, its performance significantly degrades on cross-domain datasets (SlideSpeech, GigaSpeech, etc.), revealing critical limitations in domain generalization.

\subsection{Cross-Domain Speech Recognition Text Fine-Tuning Results}
\label{sub:Effective}

As shown in Table~\ref{tab:effective}, our proposed text-only fine-tuning method achieves substantial improvements when applied to the LibriSpeech-pretrained Speech LLM (8B) model. On SlideSpeech, text fine-tuning reduces the WER by 11.37 absolute points (41\% relative error reduction), demonstrating strong domain adaptation capability. For Medical data, representing an extremely low-resource scenario (less than 1 hour of corresponding text), a WER reduction of 0.55 (4.2\% relative improvement) is observed, although the improvement is less pronounced due to the limited amount of training data. 

It is worth noting that text fine-tuning on GigaSpeech yields the best performance on the Medical domain, achieving a WER of 12.26\%. This can be attributed to the scale and coverage of GigaSpeech, which, despite its noisier text quality, provides sufficient generalization benefits for low-resource domains. However, due to the extensive size and relatively lower text cleanliness, its performance on SlideSpeech and GigaSpeech target domains is slightly inferior compared to fine-tuning directly on SlideSpeech.

Notably, SlideSpeech achieves the best overall adaptation results, not only on its own domain but also on GigaSpeech, likely because of its moderate size and higher text quality.

Furthermore, the source domain performance on LibriSpeech (Clean/Other) remains stable after fine-tuning, with no significant degradation observed. This indicates that the proposed method enhances target-domain performance while effectively preserving source-domain knowledge, mitigating catastrophic forgetting and improving cross-domain generalization. 

Overall, the WERs across different target domains converge to similar levels after fine-tuning, suggesting that the model approaches a critical point in balancing adaptation and overfitting, a phenomenon worthy of further investigation in the context of large-scale Speech LLMs. Notably, this also demonstrates the effectiveness of our real-time evaluation strategy.

\begin{figure*}[!ht]
\centering
\includegraphics[width=0.8\textwidth]{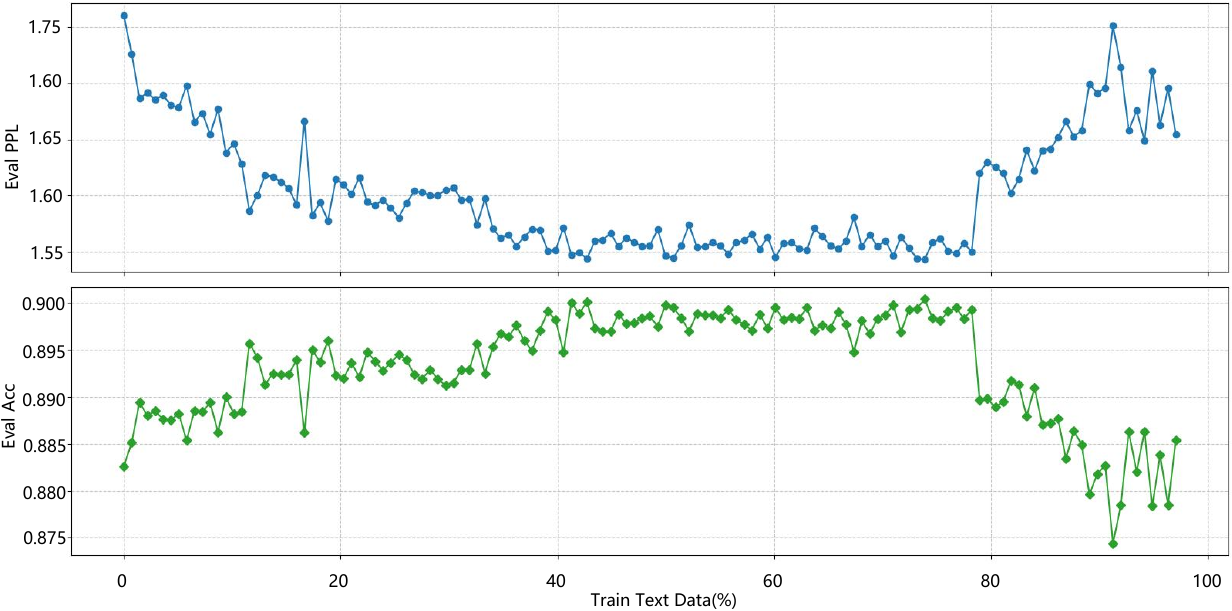}
\caption{Real-time  perplexity~(PPL) and accuracy~(Acc)  evaluation of speech alignment capabilities during text fine-tuning on the gigaspeech dataset}
\label{Fig:Eal}
\end{figure*}
\subsection{Comparison of Text Fine-Tuning and Speech Fine-Tuning}
\label{sub:comparison}

Table~\ref{tab:anaysys} presents a detailed analysis of different fine-tuning strategies under a domain adaptation setting with complete speech data. Two key observations can be made.

First, text-based fine-tuning demonstrates strong generalization to unseen domains. Although its WER on the target domain (SlideSpeech) is relatively higher than that of speech-based fine-tuning (16.22\% vs. 12.99\%), it significantly outperforms speech-based fine-tuning in the source (LibriSpeech) and other domains (Medical and GigaSpeech). Specifically, text fine-tuning results in minimal performance degradation on LibriSpeech (WER increases of only 0.63\% and 0.03\%), and achieves the lowest WERs in the Medical (13.08\%) and GigaSpeech (17.09\%) domains. This indicates that text fine-tuning effectively mitigates overfitting to the target domain and preserves robust cross-domain generalization.

Second, in terms of error types, text fine-tuning substantially reduces substitution and insertion errors, with the improvement in insertion errors attributed to better alignment induced by textual supervision. However, it slightly increases deletion errors, likely due to a weakened speech-to-text alignment capacity.

By contrast, speech-based fine-tuning achieves better WER in the target domain but suffers from substantial degradation in both source and other domains. Furthermore, it consistently reduces substitution and deletion errors, benefiting from enhanced acoustic alignment. Nevertheless, its poor cross-domain generalization limits its broader applicability.

In summary, while speech fine-tuning improves target-domain performance and alignment accuracy, text fine-tuning provides superior generalization and robust performance across domains, making it a more reliable choice in diverse deployment scenarios.

\subsection{ Real-time evaluation}

GigaSpeech has the largest amount of data and the most text materials, so we analyze the ppl, loss, and acc during the training of GigaSpeech using the real-time evaluation method mentioned above. As shown in the figure, the recognition effect of the model first increases and then decreases. The early increase is because the lora parameters of the LLM have domain adaptation to the text in this field, and the later increase is because too much text destroys the alignment ability of the lora parameters in the LLM between speech and text. The actual optimal point uses 74.6\% of the training text of the whole text, and not all the text is used. Of course, lowering the learning rate may make the time point of destroying the speech alignment later to obtain the results of all text training.

\section{Conclusion and Future Work}

This paper presents a text-based fine-tuning strategy for domain adaptation in Speech LLMs under data-scarce conditions. By leveraging only text data, the proposed method reduces the reliance on costly speech resources and mitigates the risk of overfitting to the target domain.

Experiments on multiple datasets show that text fine-tuning achieves competitive adaptation performance while preserving strong generalization across domains. A real-time evaluation strategy is further introduced to maintain speech alignment capabilities during adaptation, making the approach practical and scalable for deployment in low-resource scenarios.

However, the method still trails behind full speech-text fine-tuning in target domain performance, and the benefit is limited when LLM parameters remain frozen during pre-training. Another problem is that excessive text fine-tuning in the absence of speech can cause the model to lose its recognition ability. Although we avoid this problem by reducing the learning rate of text fine-tuning and adopting real-time evaluation, we still do not completely solve this problem. Future work will explore hybrid strategies combining text and limited speech supervision, and joint optimization of speech and language components to further enhance adaptation effectiveness.

\section{Acknowledgment}


This research was carried out by Yangui Fang and Jing Peng while they were interning at AISPEECH. We would like to express our gratitude to AISPEECH for providing the necessary computing resources to support this work.
\newpage

\bibliographystyle{style/IEEEtran}
\bibliography{main}

\end{document}